# Accelerating Scientific Research Through a Multi-LLM Framework

Joaquin Ramirez-Medina[1], Mohammadmehdi Ataei[2], Alidad Amirfazli[1]*

[1]Department of Mechanical Engineering, York University, Toronto, ON M3J 1P3, Canada

[2]Autodesk Research, Toronto, ON M5G 1M1, Canada
*Corresponding author: Email: alidad2@yorku.ca

## Abstract

The exponential growth of academic publications poses challenges for the research process, such as literature review and procedural planning. Large Language Models (LLMs) have emerged as powerful AI tools, especially when combined with additional tools and resources. Recent LLM-powered frameworks offer promising solutions for handling complex domain-specific tasks, yet their domain-specific implementation limits broader applicability. This highlights the need for LLM-integrated systems that can assist in cross-disciplinary tasks, such as streamlining the research process across science and engineering disciplines. To address this need, we introduce Artificial Research Innovator Assistant (ARIA), a four-agent, multi-LLM framework. By emulating a team of expert assistants, ARIA systematically replicates the human research workflow to autonomously search, retrieve, and filter hundreds of papers, subsequently synthesizing relevant literature into actionable research procedures. In a case study on dropwise condensation enhancement, ARIA demonstrates its capability to streamline research tasks within an hour, maintaining user oversight during execution and ultimately liberating researchers from time-intensive tasks.

## Introduction

Scientific research typically begins with a comprehensive review of existing literature, a process essential for identifying foundational theories, tracking the latest developments in the field, and pinpointing gaps in knowledge. Although indispensable, this process is often slow and methodical, made more challenging by the worldwide surge in academic publications, with global science and engineering article outputs reaching 3.3 million in 2022 (1). This explosive growth compounds an already challenging research process, particularly the literature review stage, leading to more complex and time-consuming activities for searching, retrieving, and screening relevant work. Researchers, especially those exploring unfamiliar domains, must therefore seek better ways to navigate this large, rapidly expanding volume of publications.

Large Language Models (LLMs) offer a potential solution to these challenges. LLMs are able to distill and interpret vast information, providing a more accessible output to users (2, 3). Such capabilities are especially evident in Generative Pre-Trained Transformer (GPT) models, a type of LLM powering, e.g., ChatGPT, where users can engage in detailed, contextually relevant conversations on general knowledge (4).

LLMs excel at general knowledge, which is stored in their model weights, shown by their performance in benchmarks such as General-Purpose Question Answering (GPQA) (5). GPQA is a benchmark task where models are evaluated on their ability to answer a broad range of graduate-level general questions written by experts in physics, biology, and chemistry.

Yet, these models falter when applied to niche or specialized domains. The primary reason is in the relative scarcity of specialized data in the training datasets, which mostly general content rather than highly technical or domain-specific material needed in research. This creates a gap in the general-purpose LLM's ability to precisely understand or generate content involving specialized terminology or concepts that are not frequently represented in the training data.

The challenges associated with using LLMs for specialized tasks have spurred the development of various techniques, including in-context learning (ICL), fine-tuning (FT), and retrieval-augmented generation (RAG). These techniques are designed to not only better equip LLMs for specific demands, but also to streamline the execution of nuanced, domain-specific tasks.

ICL relies on custom-designed prompts that supply LLMs with richer context, including task objectives, relevant background, terminologies, constraints, etc. Combined with a user dialogue, ICL greatly enhances the LLM’s ability to generate user-targeted, precise, and informative responses (3, 6). The FT approach, tunes pre-trained LLMs on domain-specific datasets to enhance performance on domain-specific tasks, improving their handling of complex topics (7). For example, Med-Gemini, built upon Google’s Gemini models, is a fine-tuned LLM excelling in medicine-related tasks, including text summarization, and research and educational knowledge (8). Finally, the RAG approach combines search and retrieval mechanisms with LLMs for fetching relevant data from a database of information, allowing access to real-time, domain-specific information to enrich model responses (9). RAG systems commonly employ distinct retrieval mechanisms, e.g., keyword-based, vector embedding-based retrieval or hybrid, with the latter converting text into numerical vectors to capture semantic relationships beyond simple keyword matching.

Despite the progress made through above techniques there are still limitations. For instance, the success of ICL largely depends on the precision and structuring of the manually crafted prompts, often requiring substantial prompt engineering. Additionally, their efficacy is still constrained by the model's context limit, even with recent advances. Most current models now have token limits of over 128k tokens, with some models like Gemini reaching over 2 million tokens (10). However, LLMs are charged per token, which means that using the full context and larger token capacities can become extremely costly.

FT calls for ample computational resources, an appropriate training data corpus, and access to model weights or a fine-tuning API, all of which demand significant financial investment and domain expertise. Not only are these resources often hard to obtain, but there is also a risk that fine-tuning could produce poor results, due to factors like insufficient training data, lossy compression, overfitting, or a mismatch between the fine-tuned model and its intended application. These challenges together create a substantial barrier to entry, especially for non-experts.

RAG likewise has its own difficulties, including the need for a pre-defined dataset requiring frequent maintenance and updates, which may limit adaptability for non-AI exerts. RAG frameworks require manual query formulation to access their knowledge base where important context can be missed, if queries are not well-designed, similar to the challenges in ICL. Moreover, RAG systems require the ability to handle extended contexts, particularly when processing long or complex documents that may require integration of multiple sources (11).

These challenges illustrate the broader struggle to ensure LLMs remain current, accurate, and convenient for research settings, especially for individuals lacking AI expertise.

Despite these challenges, LLMs utilizing specialized techniques have recently attracted considerable interest across various fields, including cybersecurity (12), manufacturing (13), law (14), and medicine (15). These applications leverage LLM capabilities to assimilate domain-specific knowledge, improving task precision and accelerating execution through tailored model adaptations. Among these applications, a few studies have explored the potential of LLMs to enhance systematic review processes for research and streamline complex technical tasks. The developed methods in (16) and (17), were tailored to particular academic domains and databases, using ChatGPT to generate Boolean search queries for systematic reviews. Both studies employed in-context learning through a series of guiding prompts to incrementally specialize the LLM on the review topic. While demonstrating the potential of in-context learning for nuanced content generation, these studies further highlighted the challenges associated with manual prompt crafting and the need for coupling LLM-based systems with additional resources and tools.

LLM-based frameworks introduced in (18), (19), (20), addressed complex technical tasks in building information modeling (BIM), chip design, and chemical synthesis, respectively. These domain-specific frameworks integrated LLMs within specialized modules incorporating additional resources and tools to enable autonomous prompt crafting, enhanced information retrieval, and code generation. BIMS-GPT (18) and ChipGPT (19) showcased the potential of LLM-based frameworks in reducing processing times and manual effort. Coscientist (20) demonstrated the agentic-based orchestration of multiple LLMs for streamlined experimental design and execution, utilizing vector embeddings for querying hardware documentation. However, these frameworks remained primarily accessible to specialists within their respective scientific and engineering domains, emphasizing the continuing challenge of developing LLM-based frameworks that can support researchers across different domains.

An urgent need persists for a tool that enhances research and development across various scientific and engineering fields, rather than focusing on a single domain. This tool should leverage LLM technology to enable quick, informative literature reviews, keeping pace with the accelerated growth of literature. Rather than developing a method for a single domain and/or academic database, it must provide a means to adapt to any science and engineering domain and interface autonomously with multiple academic databases. An important feature will be the ability to retrieve literature from these databases to assist users in planning their research procedures. Through this, users can gain up-to-date, actionable knowledge to inform their research design and methodologies, hence, accelerating the research process.

In this paper, we propose the Artificial Research Innovator Assistant (ARIA): Literature Review & Planning module. To this end, we employ an agent-based framework, employing specialized agents to accomplish four major objectives: 1) understanding the user's research question, 2) searching, retrieving, and screening relevant literature based on the user's research project, 3) processing and indexing screened literature into a queryable structure, and 4) synthesizing relevant literature into a suggested procedure, providing next steps for conducting the project.

To demonstrate the Literature Review & Experimental Planner module's implementation, we select an area within surface engineering that focuses on the enhancement of dropwise condensation through electrowetting and shearing airflow techniques. Achieving dropwise condensation can significantly improve heat transfer and advance cooling technologies across industries, making it an ideal testing ground for the ARIA.

## Results and Discussion

The results of development is a four-agent, multi-LLM architecture, for ARIA, see Fig. 1. When researchers begin a new project, they typically brainstorm a research question, design search queries, track down relevant studies, evaluate their importance, store full texts, take notes on key concepts, and finally propose a method or procedure based on what they learned. ARIA performs these tasks, but breaks these steps into specialized agents that work together—akin to a team of experts handing off tasks to one another. Instead of a person manually scanning for relevant articles, the Retriever Agent performs large-scale searches and filters out noise. Rather than flipping through pages of journals to identify critical methods, the Processor Agent organizes data to spotlight what matters most. Meanwhile, the Conversation Agent emulates a collaborator who asks clarifying questions and summarizes findings for easy reference. Finally, the Suggester Agent behaves like a seasoned advisor who combines all those pieces into a structured plan—pinpointing the novel elements of a study and formulating a procedure that researchers can utilize the lab. This automated approach replicates each step a human might follow, only faster and with fewer blind spots, ultimately giving researchers more time to explore ideas rather than sifting through paperwork.

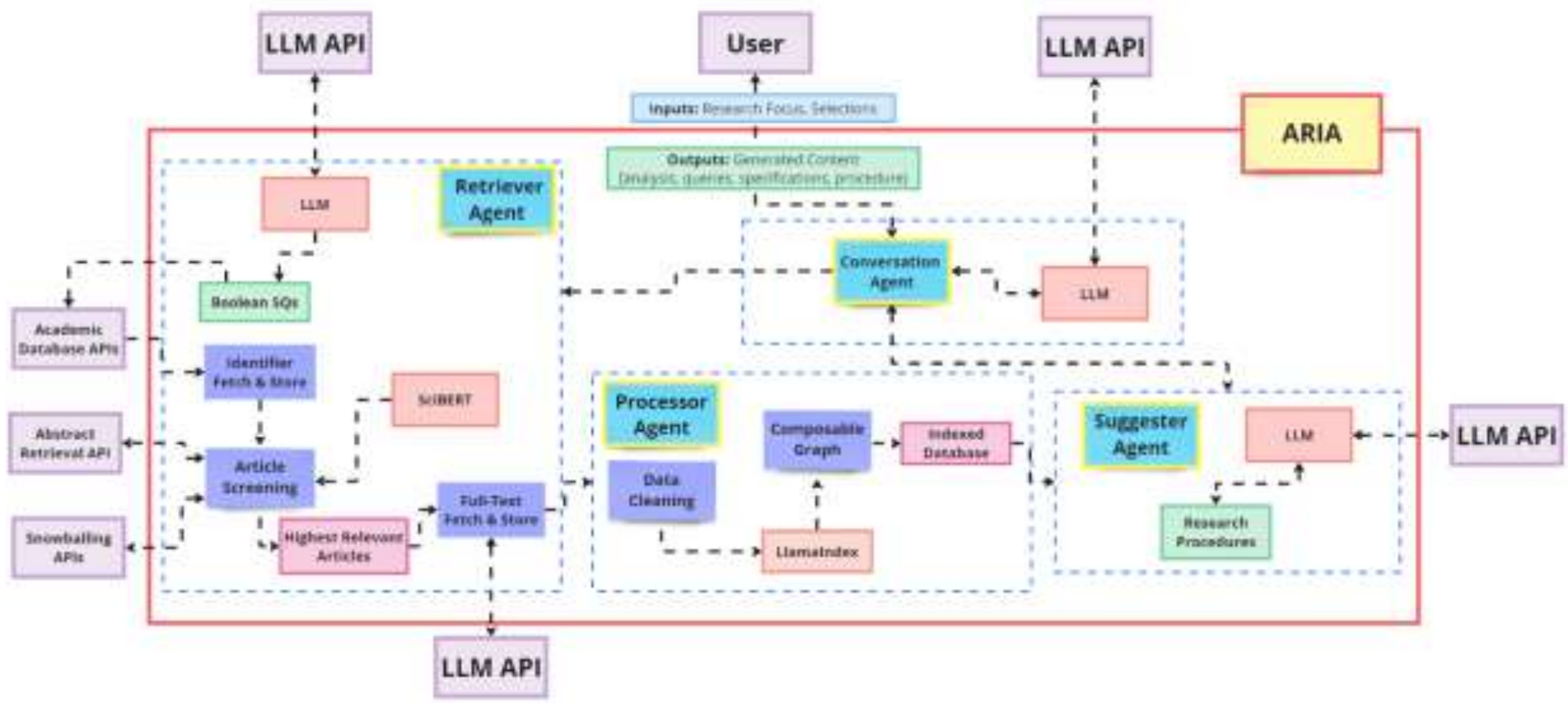

**Fig. 1. General architecture of ARIA**.

We utilize a multi-LLM, agent-based architecture for automated literature review and procedure planning, demonstrating the potential of AI-assisted tools for streamlining time-consuming research tasks. Through a case study on dropwise condensation, ARIA addresses the current challenges associated with the scientific research process, revealing two primary contributions to AI-assisted research.

First, ARIA addresses the literature review bottleneck identified across science and engineering through its automated, agent-based workflow. ARIA substantially reduces the time needed for critical literature review tasks (e.g., crafting Boolean queries, academic database querying, filtering articles for relevance, and in-depth full-text article examination). ARIA advances beyond existing LLM-based literature review frameworks in three major ways: 1) expanding the user-selected set of database queries, 2) querying three database APIs (Application Programming Interfaces) rather than focusing on a single academic database like (16) and (17), and 3) screening thousands of articles through a developed screening algorithm. Through this framework, ARIA drastically reduces literature review times to within an hour, accomplishing tasks that are usually not done or require several months.

Second, ARIA introduces a novel approach to research procedure planning. Rather than utilizing a static, pre-defined dataset, ARIA generates a dynamic, tailored dataset for the current user. Using the screened literature from previous steps and LlamaIndex tools, we create an organized hierarchical index structure that enables a user-targeted, domain specific RAG. Through this architecture, ARIA can effectively assimilate relevant technical information, allowing autonomous RAG execution without requiring AI expertise or manual user intervention. The Suggester Agent LLM then provides clear, step-by-step research procedures based on the examined literature. For the current case study, the experimental procedure comprised four key elements:

i. Specific materials, equipment, and instrumentation requirements
ii. Broader system setup and connection instructions, followed by detailed configuration guidelines for user-selected specifications
iii. Novel aspects introduced by ARIA within the detailed configuration instructions
iv. Data analysis and characterization steps

The following sections provide further details of each agent's functionality and their results through ARIA's workflow execution, demonstrating its application through the case study of dropwise condensation.

*User Research Focus Elicitation and ARIA-Generated Perspective*

Establishing a clear understanding of the user's research focus sets the stage for the ARIA execution sequence, as explained here.

The Conversation Agent provides the user a broader perspective of the user's research focus, first for confirmation by the user before diving into specific technical details. Similar to (16) and (17), the user articulates their research focus with increasing specificity to the LLM-based system through user-LLM dialogue chaining. While previous approaches used ChatGPT and its interface, ARIA's Conversation Agent serves as the user-AI interface, allowing user inputs to be utilized throughout its multi-agent system. This provides ARIA with a focused research context for better understanding the objectives and key areas of inquiry before taking next steps. For the current case study, Fig. 2 (A-C) illustrates the refinement of a user's research focus from an initial broad topic to specific technical components.

**A. Research Topic:** What is the general topic you will be delving into for phrasing your topics today?

Below are some examples that you may use as a guideline.

- 1. Separation and purification of bio-based acetone butanol ethanol production
- 2. Acoustic-based rotational manipulation of drosophila embryos

**User input topic:** *Dropwise condensation in surface engineering*

---

**B. Research Objectives:** Within your topic, what are your overarching objectives?

Below are some examples you may use as a guideline.

- 1. Developing a purification system for acetone butanol ethanol production via numerical means
- 2. To develop a high throughput trapping and manipulation system for real-time imaging

**User input objectives:** *Enhancing dropwise condensation, heat transfer transfer/rate enhancement*

---

**C. Research Aspects:** What specific aspects of your topic will you be investigating to achieve your objective(s)? Consider specific components within these aspects that you wish to further examine in closer detail.

Below are some examples you may use as a guideline.

- 1. GSVP, conventional distillation columns, vacuum batch fermentation, pervaporation membrane
- 2. Piezoelectric transducers for microstreaming, PDMS-based system, air bubble microcavities

**User input aspects:** *Droplet shedding by electrowetting and shearing airflow, electrode design and arrangement*

---

**D. Analysis**
**Assistant:**
Dropwise condensation refers to the formation and growth of individual droplets on a surface during the condensation process. It is an important phenomenon in many industrial applications, as it can significantly enhance heat transfer rates compared to filmwise condensation, where a continuous condensate film forms on the surface.

One aspect of dropwise condensation is droplet shedding, which refers to the removal of condensed droplets from the surface to maintain a clean and active condensing surface. There are two main methods for droplet shedding: electrowetting and airflow.

Electrowetting involves applying an electric field across the condensing surface, which alters the wettability of the surface and promotes droplet shedding. By applying a voltage to the surface, the contact angle between the droplet and surface can be changed, allowing for easier shedding.

Airflow, on the other hand, utilizes the force exerted by a gas or air stream flowing over the surface to remove the droplets. This method is commonly used in cooling systems, where a continuous flow of air helps to sweep away the droplets.

Another important aspect of dropwise condensation is heat transfer enhancement. Dropwise condensation provides higher heat transfer coefficients compared to filmwise condensation due to the increased surface area and efficient droplet shedding. This enhancement is crucial in various applications, including power generation, heat exchangers, and cooling systems.

The electrode arrangement and design play a significant role in facilitating droplet shedding through electrowetting. The arrangement of electrodes on the condensing surface determines the efficiency and effectiveness of the shedding process. Optimizing the design can ensure uniform shedding of droplets and enhance the overall heat transfer performance.

These different aspects of dropwise condensation are interconnected. The effectiveness of droplet shedding methods such as electrowetting and airflow depends on the surface properties and wettability. A well-designed electrode arrangement can enhance the shedding process and ultimately improve heat transfer efficiency. The combination of these aspects leads to more efficient and effective dropwise condensation, enabling better heat transfer performance in various applications.

**Fig. 2. User research focus definition and crafted analysis for the case study.** (**A**) General research topic input. (**B**) Research objectives input, demonstrating how to articulate specific goals. (**C**) Research aspects of interest, illustrating the breakdown of technical components. This framework guides researchers in progressively refining their research focus from broad topic to specific technical elements. (**D**) Resultant generated analysis enriching the initial research focus, illustrating the fundamental principles within the first six paragraphs, and the key interconnections within the final paragraph.

The final response in Fig. 2D is generated by systematic analysis of two key elements: fundamental principles and aspect interconnections. For the fundamental principles, the LLM identifies the underlying key concepts and provides concise definitions on each. Additionally, the LLM explores how these aspects interact to influence outcomes and accomplish the overarching objectives.

If unsatisfied, the user can input missed aspects or request a new analysis. Any additional aspects the user wishes to investigate are incorporated into the newly crafted analysis.

Figure S1 of the Supplementary Materials visualizes the execution sequence for the first Conversation Agent function for crafting this analysis, and to re-generate the analysis, if desired. Our strategy leverages the LLM's inherent capabilities through guided in-context learning prompts, rather than modifying its internal weights or architecture. Figure S2, details how the Conversation Agent LLM is programmed via in-context learning instruction.

Database suitable queries are developed through a systematic process that begins with formulating search phrases in natural language. This preliminary creation of phrases ensures a consistent natural language foundation before applying database-specific Boolean operators and syntax structures.

For the current case study, Fig. 3A displays the Conversation Agent-crafted search phrases. If the generated phrases spark new interests, users can input custom phrases, which are then appended to the list of previously selected phrases, as depicted in Fig. 3B.

To craft the targeted phrases, the Conversation Agent LLM is instructed to take a “bottom-up” approach. In this approach, keyword units – i.e., the building blocks for phrase construction– are discovered by the LLM and systematically organized into three thematic categories:

- *Method Theme*: Encompasses core mechanisms, techniques, or approaches used to investigate or manipulate the subject of study (e.g., measurement techniques, analytical procedures, experimental protocols, computational approaches, processing techniques).
- *System Theme*: Comprises key elements, materials, or parameters being studied or controlled (e.g., materials, components, parameters, variables, structures, configurations, conditions).
- *Target Theme*: Includes intended outcomes or applications being pursued (e.g., applications, functions, outcomes, performance metrics, desired properties).

See fig. S3 for a depiction of how the LLM conducts the bottom-up approach through ICL.

**A. Conversation Agent Crafted Search Phrases**

- "Improving dropwise condensation efficiency with hydrophobic surface coatings"
- "Electrowetting techniques for enhanced droplet shedding in condensation"
- "Role of shearing airflow in droplet shedding during dropwise condensation"
- "Design and arrangement of electrodes in electrowetting for condensation applications"
- "Optimization of electrode shape and geometry for efficient droplet shedding in electrowetting applications"
- "Durability and performance of superhydrophobic coatings in condensation"
- "Material properties influencing droplet shedding in condensation systems"
- "Thermal conductivity and heat flux in dropwise condensation"
- "Enhancing droplet adhesion and shedding through electrode design"
- "The effects of electrowetting on heat transfer and nucleation during dropwise condensation"

**B. User Phrase Selections and Custom Phrase Inputs**

- "Electrowetting techniques for enhanced droplet shedding in condensation"
- "Design and arrangement of electrodes in electrowetting for condensation applications"
- "Material properties influencing droplet shedding in condensation systems"
- "Thermal conductivity and heat flux in dropwise condensation"
- "The influence of contact angle and surface tension on droplet shedding in electrowetting applications"

**Fig. 3. Conversation Agent crafted search phrase for the case study.** (**A**) Displays the LLM crafted search phrases for the case study, with the user selected phrases in blue. (**B**) Shows the final user selected search phrases with the custom user input search phrase (red) appended to the original phrase list.

### *Article Data Retrieval and Screening*

To maximize relevant article retrieval across academic database searches, the Retriever Agent conducts a search phrase expansion. This prevents omitting potentially relevant papers that express similar concepts using different terminology. For the current case study, fig. S4 and fig. S5 display the result of this expansion and how this process is carried out, respectively. A Retriever Agent LLM is employed to "expand" each search phrase into a morphological equivalent or variant on the premise of keyword units and themes. For the cases where users input custom search phrases, the LLM decomposes the phrases to reconstruct keyword units and themes and then proceeds in expansion.

Once the expansion is completed a Retriever Agent LLM transforms each phrase into Boolean queries for academic database retrieval. To achieve high precision and recall, the LLM is instructed to break apart each search phrase into keyword units, and then reconstruct themes utilizing Boolean operators (e.g., AND, OR) as keyword unit connectors. Figure 4 depicts the result of this transformation for the case study. See fig. S6 for details of how the transformation process is achieved through ICL prompting.

**Boolean Query Transformation**

- Search Phrase 1: "electrowetting techniques AND (enhanced droplet shedding OR droplet removal) AND condensation"
- Lateral Expansion 1: "Optimization AND electrowetting AND ("droplet shedding efficiency" OR "droplet removal") AND condensation systems"

- Search Phrase 2: "electrodes AND (design OR arrangement) AND electrowetting AND condensation applications"
- Lateral Expansion 2: "Strategic positioning AND conductive elements AND electrowetting AND ("droplet shedding efficiency" OR "droplet removal") AND condensation processes"

- Search Phrase 3: "properties AND material AND (influencing OR affecting) AND droplet shedding AND condensation"
- Lateral Expansion 3: "Impact AND surface characteristics AND ("liquid droplet removal" OR "droplet shedding efficiency") AND condensation setups"

- Search Phrase 4: "thermal conductivity AND (heat flux OR heat transfer"
- Lateral Expansion 4: "Analysis AND heat transfer rates AND thermal properties AND ("dropwise condensation mechanisms" OR "dropwise condensation process")"

- Search Phrase 5: "contact angle OR surface tension) AND droplet shedding AND electrowetting"
- Lateral Expansion 5: "Relationship AND liquid-solid interactions AND surface properties AND ("droplet removal efficiency" OR "droplet shedding efficiency") AND electrowetting systems"

**Fig. 4. Transformation result of search phrases into Boolean queries for the case study.**

Unlike the previous studies (16) and (17), the Retriever Agent routes the Boolean queries to multiple academic database API (e.g., ScienceDirect, Web of Science, and Scopus) for article identifier retrieval, returning either the Digital Object Identifiers (DOIs) or titles for each article. For the current case study, the ten finalized Boolean queries returned 313 articles identifiers (i.e., 313 academic paper article DOIs and titles) from the databases within a matter of minutes. By simultaneously routing each query to all three APIs, corresponding lists of relevant identifiers are returned.

To broaden the literature search, without overlooking any potentially relevant papers that may not be retrieved using the above technique, an article reference and citation tracking or "snowballing" is employed. For the case study, the Semantic Scholar API is employed for snowballing, which returned an additional 1,314 papers in addition to the previously retrieved 313 articles, increasing the total article count to 1,627 papers across all API bins. Snowballed article identifiers are then appended to their respective source API bin.

To rapidly filter articles that are most relevant to the user's research, a frequency-based ranking stage is employed across the API bins. Based on a fixed threshold, frequently appearing articles are placed in a Final List, where less frequent articles are assigned to a Low Ranked bin. For the case study, 29 high frequency articles are identified and stored in the Final List bin. The remaining 1,598 low frequency articles are stored within the Low Ranked bin.

To avoid prematurely excluding articles that appear less frequently across databases, but may still be significant to the user's research, a semantic-based screening via SciBERT is conducted utilizing vector embeddings between abstracts and search phrases. In the case study, this method identified 53 additional relevant articles out of the 1,598 total low frequency articles, increasing the total count of the Final List from 29 to 82 article identifiers. To accomplish this, the Retriever Agent fetches the full text abstracts of the low-ranked articles through two APIs, including the Elsevier's Abstract Retrieval API and the arXiv API. The former leads the DOI-based abstract retrieval, whereas the latter targets title-based abstract retrieval. Additionally, the Retriever Agent fetches the finalized search phrases from the Conversation Agent. Subsequently, abstracts and phrases are converted to vector representation and their cosine similarities are calculated, as illustrated in fig. S9. Abstracts passing a fixed semantic similarity threshold have their article identifier appended to the Final List.

In constructing the Final List bin– representing the top relevant article (as article identifiers) for the user's research focus– the Retriever Agent begins full text retrieval of each article through two APIs, including the Elsevier Full-Text Retrieval API and the arXiv API. Utilizing the Elsevier API provides comprehensive coverage of article DOIs, whereas the arXiv API provides access to emerging works (i.e., preprints), expanding the collection beyond the Elsevier database. For the current study, out of the 82 articles present on the Final List, approximately 28 full text papers were returned. The Elsevier API returns articles in plain text format and are stored in an Elsevier Full Text bin, while arXiv provides articles as PDFs and are stored in an arXiv PDF bin. See fig. S10 for the results and illustration of full text retrieval process, additionally illustrating the full text storage according to the API source format. The above strategy not only speeds up the retrieval and analysis of papers, but is also economical as it substantially reduces token traffic.

*Preprocessing and Index Configuration*

To ensure refined, uniform data across retrieved full texts, facilitating a standardized cleaning and indexing process, all arXiv returned PDF articles are subsequently converted to plain text format. A Python PDF-to-text extraction library is utilized to extract and store article plain texts within an arXiv Full Text bin. The Processor Agent inspects each article within the full text bins and removes any superfluous metadata (e.g., metadata, tags, etc.) leaving only the essential literature content. Once completed, the agent consolidates all cleaned article texts from both Full Text bins into a central Index Data bin, as illustrated in fig. S11.

Similar to Coscientist (20), we employ a vector embedding-based retrieval to query information from the returned literature. However, to efficiently manage the search, retrieval, and synthesis of large volumes of scientific texts while still maintaining retrieval precision, we employ the data management framework, LlamaIndex. Through LlamaIndex, we are provided a means for seamless integration of multiple complex documents through its hierarchical composable graph index structure. The index structure enables both broad and granular RAG operations through dedicated query engines that recursively traverse from high-level document summaries to detailed content representations. This allows for efficient traversal of technical, domain-specific information and subsequent synthesis, whether for dropwise condensation or any other scientific engineering domain of interest. Moreover, combining ARIA's literature retrieval capabilities with the data management framework, we can create a flexible knowledge structure that adapts each user's research beyond the single-domain constraints in (18-20).

The information is then returned to the master query engine LLM to process, synthesize, and respond to queries based on the retrieved literature context.

*Information Synthesis and Research Generation*

To develop a precise research procedure that addresses the specific elements of the user's investigation, the Conversation Agent is re-engaged. User specifications of interest are retrieved through a user-agent dialogue sequence where a Conversation Agent LLM is employed to generate a list of key specifications that the researcher should pay attention to in the project of interest. For the current case study, Fig. 5 displays the generated specifications, ranging from characterization procedures to measurement methods. The user is also provided with the opportunity to input additional specifications of their choosing.

**Key Specifications/Parameters**

- "Hydrophobic surface preparation"
- "Surface roughness quantification method"
- "Hydrophobicity/hydrophilicity characterization"
- "Electrode pattern design"
- "Airflow control method"
- "Dielectric layer optimization process"
- "Electrode material selection"
- "Droplet generation method"
- "Droplet volume control method"
- "Droplet movement tracking"
- "Coalescence and splitting observation"
- "High-speed imaging setup for droplet dynamics"
- "Surface tension measurement"
- "Droplet evaporation rate measurement"
- "Voltage application and control system"
- "Electrowetting frequency modulation system"
- "Voltage application and control system"
- "Humidity control mechanism"
- "Temperature regulation system"
- "Droplet nucleation rate measurement"
- "Heat transfer measurement method"
- "Contact angle measuremen"
- "Droplet shedding rate measurement"

**Fig. 5. ARIA crafted specifications**. Displays the user-selected specifications in blue.

To craft research specifications, the Conversation Agent LLM leverages the previous crafted analysis, using the definitions of themes from the phrase crafting section, to identify and return the most relevant Method and System Theme specifications. See fig. S13 for a depiction of the process through ICL.

The Suggester Agent is then engaged to begin the final stages for research procedure generation. The Suggester Agent begins by consolidating all user research focus inputs and selections into an organized blueprint, providing ARIA with a centralized source of user research information, including research focus (i.e., topic, objectives, aspects, and the selected specifications and parameters). To create this blueprint, a Python dictionary structure is utilized, populating research information headings as keys with their corresponding data as values.

To autonomously initiate the vector embedding-based retrieval of information we utilize an input query, similar to Coscientist (20). However, to accommodate users across different scientific and engineering domains, the Suggester Agent dynamically generates a customized in-context learning prompt leveraging the current user's research focus. This prompt is then used to execute RAG-based retrieval on the composable graph index while also providing the master query engine three main instructions on how to query the index:

1. Utilize the consolidated blueprint as the contextual basis for search and retrieval.
2. Identify areas within the index that have potential for novelty and innovation.
3. Utilize the contextual basis and the relevant retrieval information to craft a research procedure incorporating the identified novel aspect.

Figure. S14 provides an illustration of this process.

The master query engine navigates the graph, searching and retrieving information for addressing the in-context learning instructions. Utilizing a built-in LLM (i.e., OpenAI GPT model: gpt-4), the master query engine synthesizes the information returned from the graph to return a research procedure, as displayed in Fig. 6 for the current case study. If the user is satisfied, the process concludes. If needed, the user can request a regeneration of the research procedure, with particular emphasis as well, e.g., further elaboration on the range of parameters/specifications relevant to the study, see fig. S15.

**Crafted Research Procedure**
**Assistant:**
**1. Materials and Equipment:**
a. Applied voltage source (Keithley 2450 SourceMeter)
b. Electrode shape (Circular, Rectangular)
c. Electrode material (Gold, Silver)
d. Heat transfer coefficient meter (TC-36-25-RS485, Omega)
e. Thermal conductivity meter (KD2 Pro, Decagon)
f. Droplet generator (DG-1000, Microdrop)
g. Airflow velocity device (TSI 9555-X)
h. Contact angle goniometer (Theta, Biolin Scientific)
i. High-speed camera (Phantom v2640, Vision Research)
j. Microscope (Leica DMi8)
k. Data acquisition system (DAQ) (34970A, Keysight)

**2. Experimental Setup:**
a. Connect the applied voltage source to the selected electrode shape.
b. Place the chosen electrode material on the electrode shape.
c. Connect the heat transfer coefficient meter to the thermal conductivity meter.
d. Position the droplet of the desired size on the droplet generator.
e. Connect the airflow velocity device to the contact angle goniometer.
f. Set up the high-speed camera and microscope to observe the droplet shedding process.
g. Connect the data acquisition system to the camera and microscope.

i) Droplet Generation and Characterization:
a. Generate a droplet of the desired size using the droplet generator.
b. Measure the contact angle of the droplet using the contact angle goniometer.
c. Record the droplet shedding process using the high-speed camera and microscope.
d. Analyze the droplet shedding process using the recorded data.

ii) Heat Transfer and Thermal Conductivity Measurements:
a. Measure the heat transfer coefficient of the electrode using the heat transfer coefficient meter.
b. Measure the thermal conductivity of the electrode using the thermal conductivity meter.
c. Record and analyze the heat transfer and thermal conductivity data using the data acquisition system.
iii). Electrode Design and Geometry:
a. Choose the desired electrode shape (circular or rectangular).
b. Connect the applied voltage source to the electrode.
c. Adjust the electrode shape, material, and size to the desired parameters.
d. Measure the heat transfer coefficient and thermal conductivity of the electrode.
e. Measure the droplet size.

iv) Novel Aspect - Electrode Material:
a. Choose the desired electrode material (gold or silver).
b. Apply the electrode material to the electrode shape.
c. Measure the heat transfer coefficient and thermal conductivity of the electrode material.
d. Record and analyze the heat transfer and thermal conductivity data using the data acquisition system.

**3. Data Analysis:**
a. Analyze the droplet shedding process using the recorded data.
b. Calculate the airflow velocity using the droplet shedding process data.
c. Analyze the effect of the electrode shape, material, and size on the droplet shedding process.
d. Analyze the effect of the heat transfer coefficient and thermal conductivity on the droplet shedding process.
e. Analyze the effect of the droplet size on the droplet shedding process.
f. Compare the results with the theoretical models.
g. Draw conclusions from the results.

**4. Characterization:**
a. Characterize the electrode material and droplet using the microscope.
b. Capture the droplet shedding process in real-time using the high-speed camera.
c. Analyze the images and videos using the data acquisition system.
d. Compare the experimental results with theoretical predictions and draw conclusions

**Fig. 6. The results of the Suggester Agent-crafted procedure for the current case study.**

## Methods

An overview of ARIA's architecture is provided in Fig. 1, following subsections providing a brief description of the methods of how each agent was developed. When LLMs are referenced, an OpenAI GPT model, gpt-3.5-turbo, was utilized unless stated otherwise.

### *Conversation Agent*

The Conversation Agent is the primary interface for user-ARIA interactions. It drives ARIA's execution through four functions as depicted in Fig. 7: (1) retrieving the user's research focus, (2) search phrase creation, (3) research tools and parameter specifications creation, and (4) research procedure actions.

For the first function, the exchange includes users elucidating their research focus through guided text input fields for their general research topic, objectives, and key aspects of interest. A Conversation Agent LLM processes and analyzes the provided research focus to generate a comprehension analysis. The crafted analysis serves as the context basis for all subsequent ARIA processes.

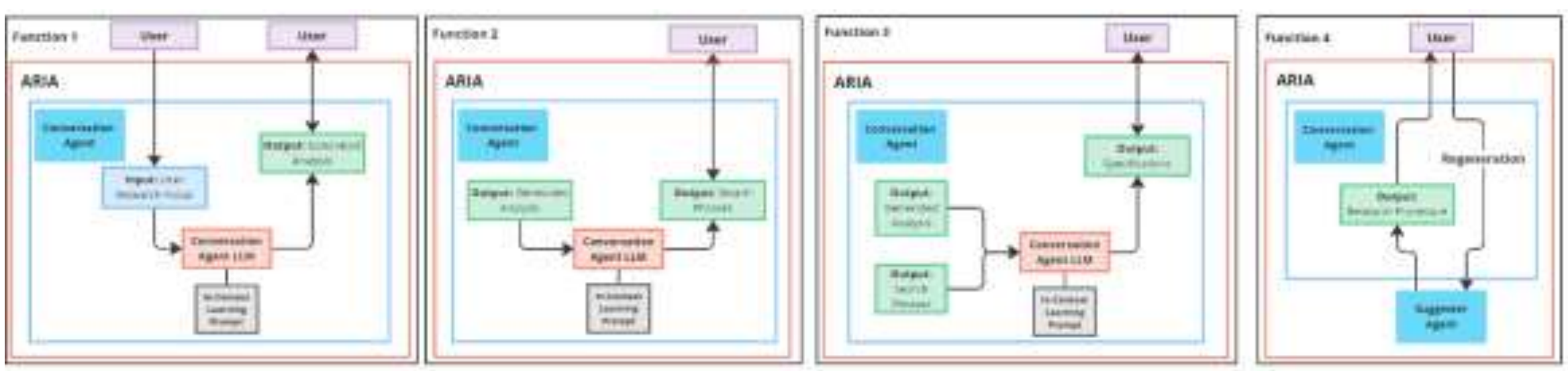

**Fig. 7. ARIA's Conversation Agent Architecture:** Displays the Conversation Agent's principal function driving the application's execution.

For function (2), ARIA leverages the crafted analysis from the first stage to provide a list of natural language search phrases capturing the essence of the user's key research inquiry. The crafting of these phrases serves as a first step for Boolean query creation suitable for academic databases. Boolean queries are used later by ARIA to search academic databases using APIs to return relevant literature.

The third function (3) of the Conversation Agent is to provide users with a range of research specifications and parameters as a list, at the end of ARIA's retrieval and analysis of literature.

The fourth function (4) is also used at the end of ARIAs workflow, to deliver the generated research procedure to the user, allowing for subsequent user actions, including a re-generation, or simply conducting the work.

### *Retriever Agent*

This agent's principal objective is retrieving article data and screening, through three main tasks: (1) article identifier retrieval, (2) article screening, and (3) full-text retrieval. See Fig. 8 for an overview of the agent's workflow.

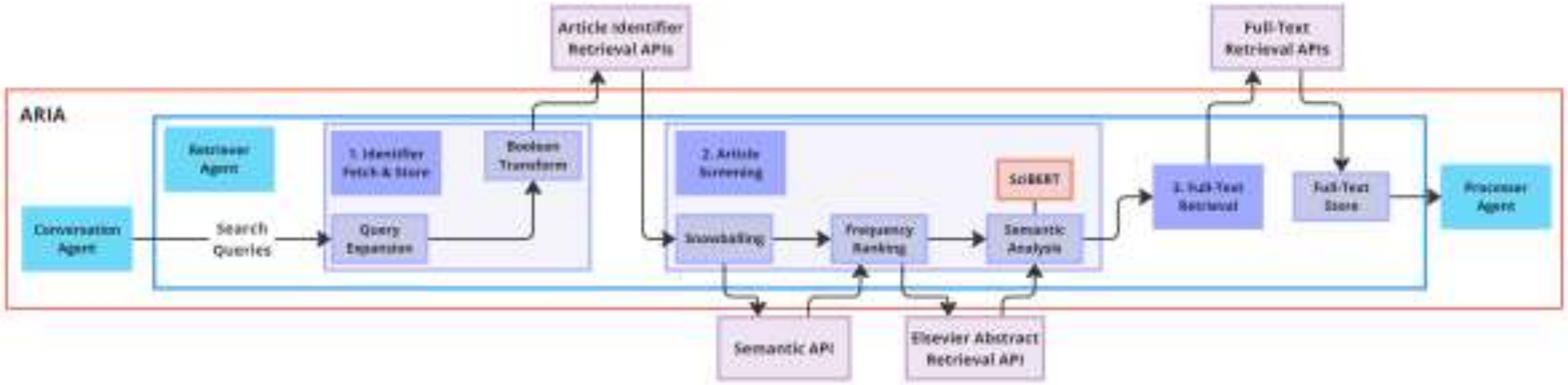

**Fig. 8. ARIA's Retriever Agent Architecture**. Sequential pipeline for academic article retrieval, screening, and processing, from initial search queries to full-text retrieval.

The article identifier retrieval task includes two main subtasks using LLMs: query expansion and subsequent Boolean syntax-based transformation. Query expansion employs an LLM to generate variations of the search phrases, resulting in a finalized list of phrases. Then the LLM transforms these finalized phrases into Boolean queries to be routed to the article identifier retrieval APIs (i.e., ScienceDirect, Web of Science, and Scopus). The primary article identifiers retrieved by ARIA include DOIs (Digital Object Identifiers) and article titles.

To screen articles, ARIA executes a three-stage algorithm: a snowballing screening stage, an article frequency ranking stage, and a semantic screening stage. In the snowballing stage, forward and backward snowballing tracks all returned article identifiers for citations and references, respectively. Next, the ranking stage employs a cross-referencing frequency ranking procedure to refine the volume of retrieved articles. Retrieval of high occurring article abstracts is conducted utilizing the Elsevier Abstract Retrieval and arXiv APIs, whereas low occurring articles are discarded for the moment. Lastly, the semantic-analysis stage employs a SciBERT model, (a fine-tuned model trained on a corpus of scientific text) for in-depth article abstract screening based on semantic similarity to the user research focus.

See fig. S7A for a detailed illustration of how the identifiers are retrieved utilizing one of the user's selected Boolean queries as an example; fig. S7B gives details of how the Article Identifier Store bin is populated with the returned identifier lists. Figure S8 illustrates the employed snowballing process for forward and backward reference tracking.

*Processer Agent*

After the retrieval and storage of full text articles, the Processor Agent conducts two main tasks as illustrated in Fig. 9: (1) data cleaning and (2) data indexing.

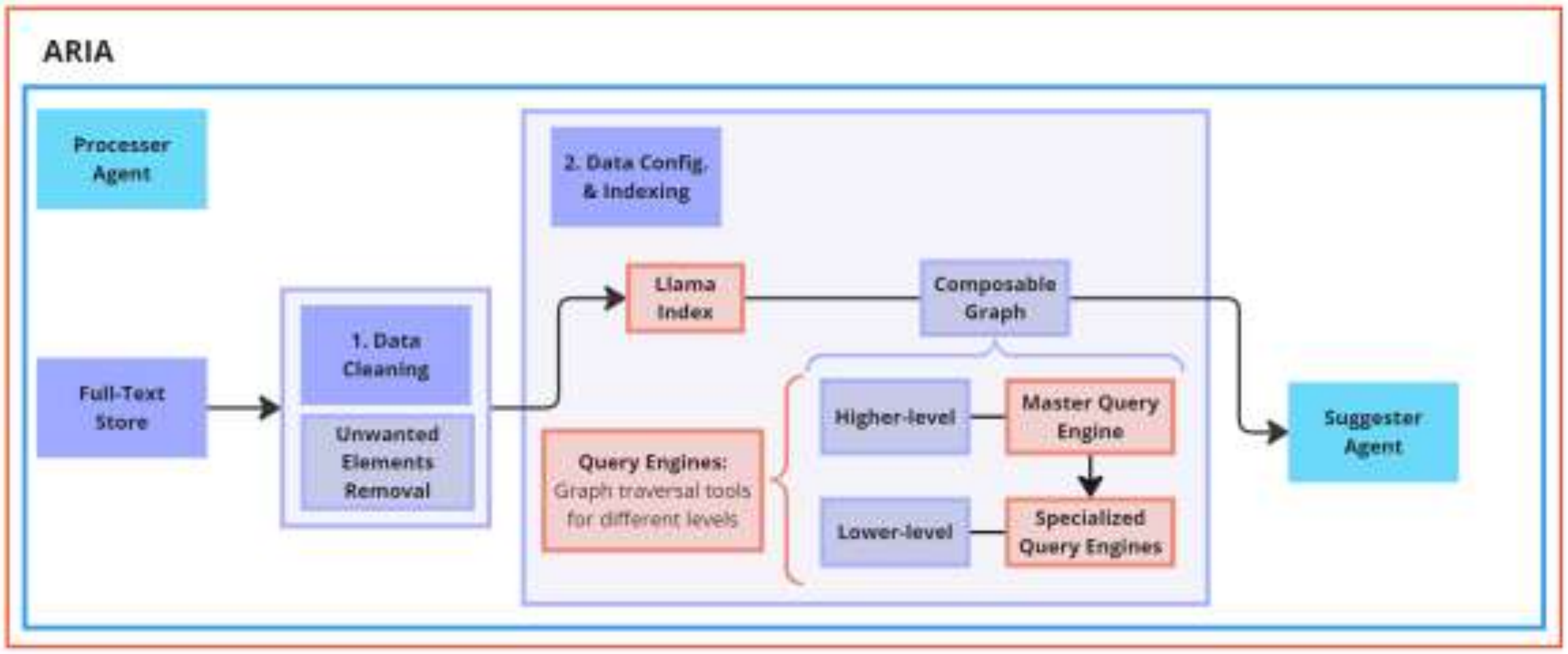

**Fig. 9. ARIA's Processor Agent Architecture**. System flow for post-retrieval article processing, showing data cleaning, as well as configuration and hierarchical indexing stages using LlamaIndex framework for RAG-based querying.

In the data cleaning stage, the Processor Agent removes unwanted elements (e.g., metadata, tags, etc.) to optimize storage and ensure uniform content across all stored articles. For quick and efficient querying (i.e., search and retrieval) of technical, domain-specific information, a data management framework, LlamaIndex, is employed. Through LlamaIndex, a queryable, RAG-based composable graph index is crafted on the literature. The composable graph provides a structured hierarchical representation for the literature present in two different formats: summary representations and detailed representations. Each screened article is assigned a single summary representation and a single corresponding detailed representation. The summary representations serve as higher-level abstractions (synopses) of each article, while the detailed representations encode the entire content of each article into vector embeddings. To achieve this, the Processor Agent leverages LlamaIndex, ComposableGraph tool.

To retrieve summaries for each article in the Graph Data bin, the Elsevier Abstract Retrieval API is employed, where LlamaIndex, ListIndex organizes the summaries for each article. For the detailed representations of each screened article, LlamaIndex, VectorStoreIndex is employed to create vector embedded representations for each. See fig. S12 for a clear illustration of this process.

Furthermore, LlamaIndex provides a query engine tool that uses the generated vector embeddings to perform RAG-based retrieval on the index. This tool functions both as a master query engine, and as a custom query engine. When a query, (e.g., a question) is provided to the graph, a two-step sequence is executed by the master query engine:

1. The engine traverses through the ListIndex containing summary representations to identify the most relevant articles addressing the input query.
2. For each relevant article identified from the summary level, the engine makes a recursive call to query that article's detailed vector index representation, extracting precise information to address the input inquiry

*Suggester Agent*

The Suggester Agent is responsible for four main tasks, as illustrated in Fig. 10: 1) creating a research blueprint, 2) final instruction prompt crafting, 3) graph querying, and 4) procedure synthesis.

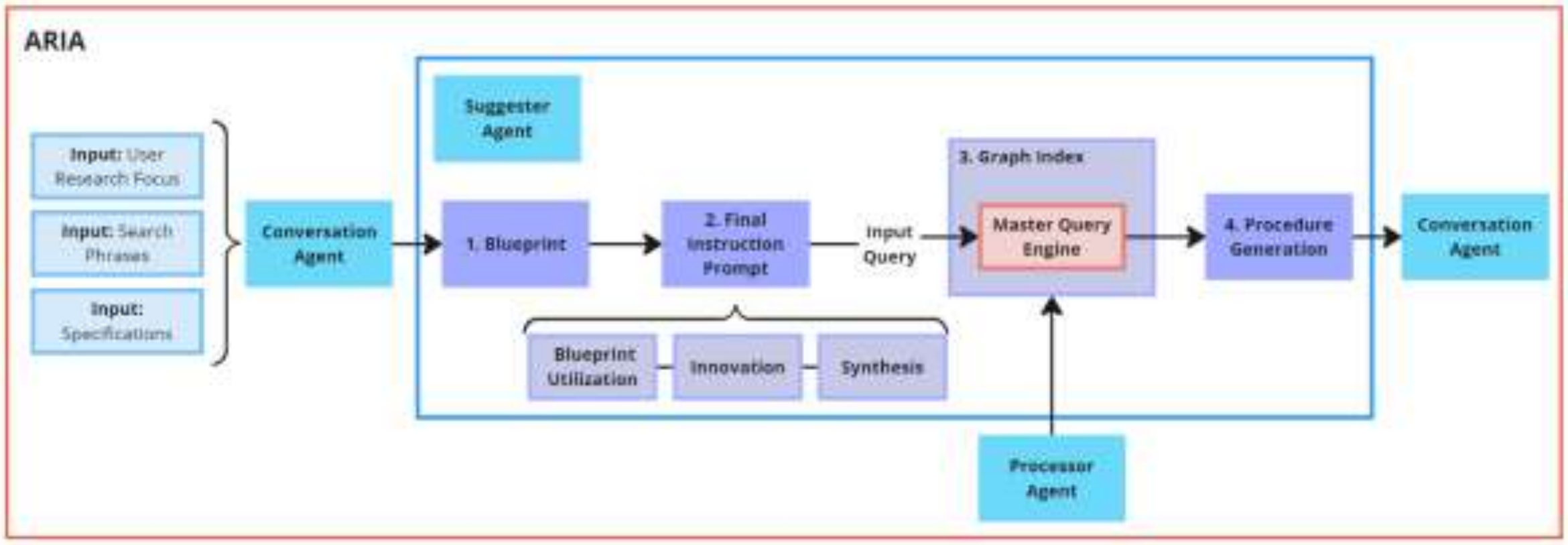

**Fig. 10. ARIA's Suggester Agent Architecture**. Workflow depicting the progression of user research inputs into a research procedure through blueprint creation, instruction prompting, graph querying, and procedure synthesis.

The user-specific research elements accumulated up to this point are consolidated by the Suggester Agent into a blueprint serving as a centralized source of information. A specialized instruction prompt is crafted by the Suggester Agent acting as the input query to guide the master query engine for the RAG process. This prompt incorporates directions for: utilizing the blueprint context, identifying innovative approaches, and synthesizing retrieved information into a final procedure. The query engine traverses the graph hierarchy, searching and retrieving relevant technical information that addresses the user's research inquiry through the blueprint. Drawing from the retrieved graph information the master query engine LLM synthesizes comprehensive research tools, parameters of interest and their ranges, and procedures.

## Limitations

However, there is room for improvements. One notable limitation is evident from ARIA's current implementation of APIs for identifier and full-text literature access. Expanding the current article identifier retrieval APIs would enable ARIA to cast a wider initial search net. Furthermore, while ARIA successfully identified 82 highly relevant papers through its screening process, only 28 full texts were successfully retrieved through the Elsevier and arXiv APIs. Expanding full-text access to include additional publisher APIs can increase literature coverage. These improvements would provide users with a more a comprehensive literature base for research planning, which is particularly valuable for interdisciplinary research topics where relevant papers are often distributed across multiple publishers' databases.

By selectively extracting, preprocessing, and indexing methodology-specific content, we can potentially reduce computational and monetary costs, while improving information retrieval relevance. This targeted approach can enhance research procedure generation by focusing RAG operations on the most procedure-relevant portions of the literature.

While full text article processing can help provide a foundational understanding of the user's research project, significant information lies within equations and graphics. Future tasks can include incorporating data from equations and graphics within the ARIA system. Specifically, integrating efficient equation and image extraction resources such as optical character recognition, GPT Vision, and PDF content extraction tools can enhance ARIA's understanding and generation of domain-specific procedures. A robust multi-modal paradigm that organizes diverse data types and maintains RAG provenance would enable ARIA to generate more comprehensive, detailed, and precise research procedures.

## Supplementary Materials

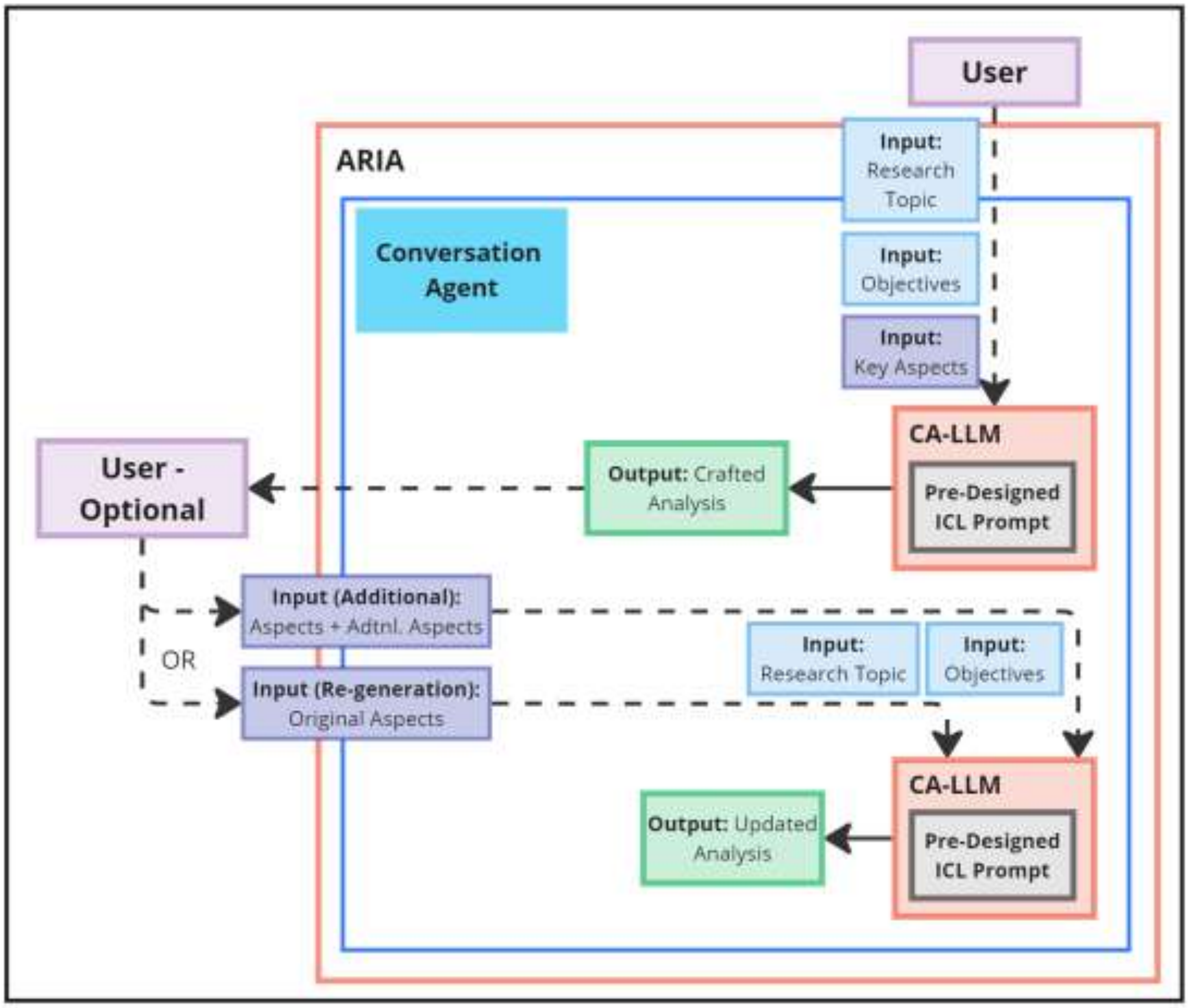

**Fig. S1. Conversation Agent detailed workflow of first function.**

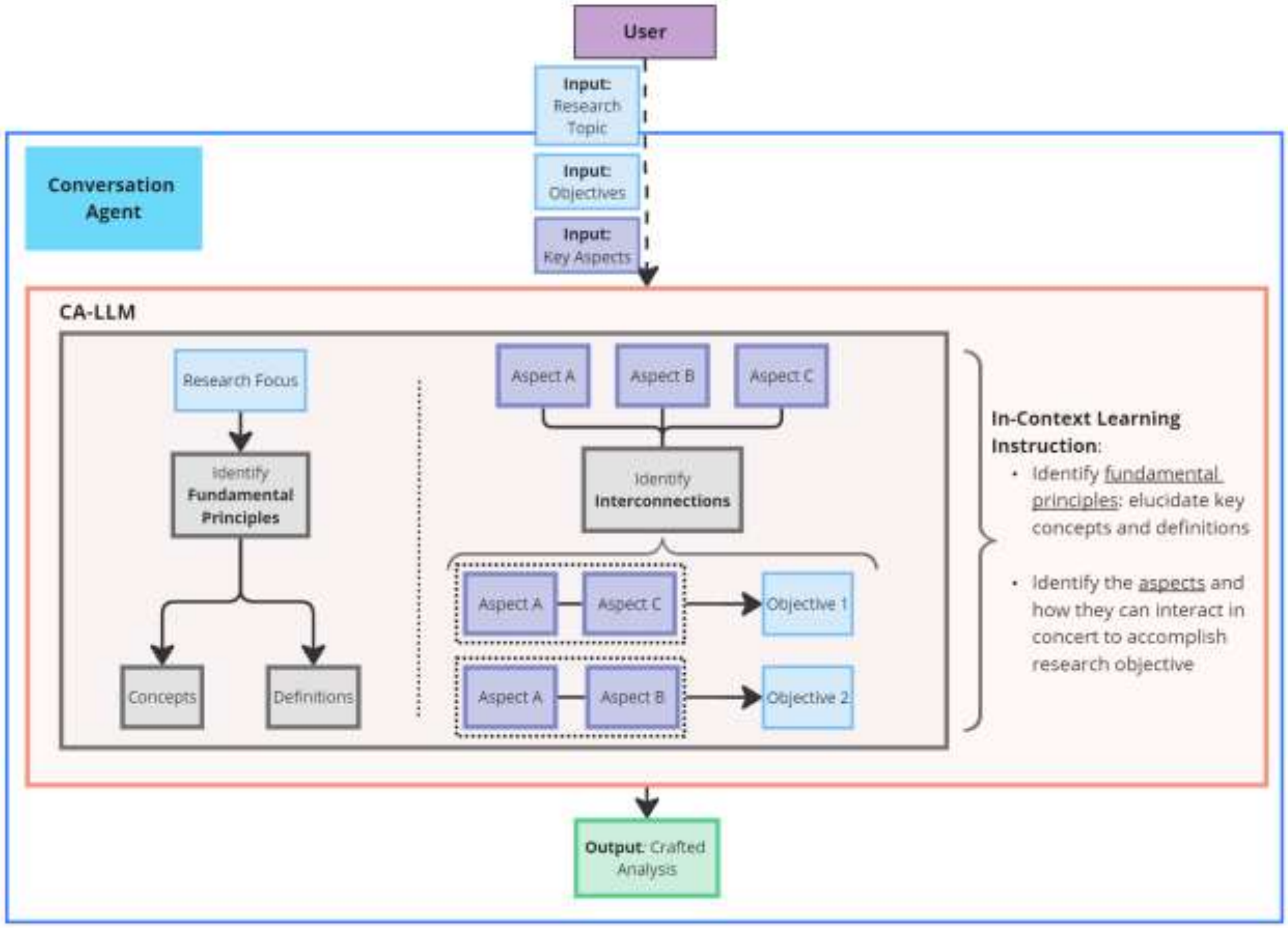

**Fig. S2. Conversation Agent LLM programming via in-context learning instruction.** Displays the process for synthesizing a broad, comprehensive analysis through fundamental principle identification and aspect interaction mapping.

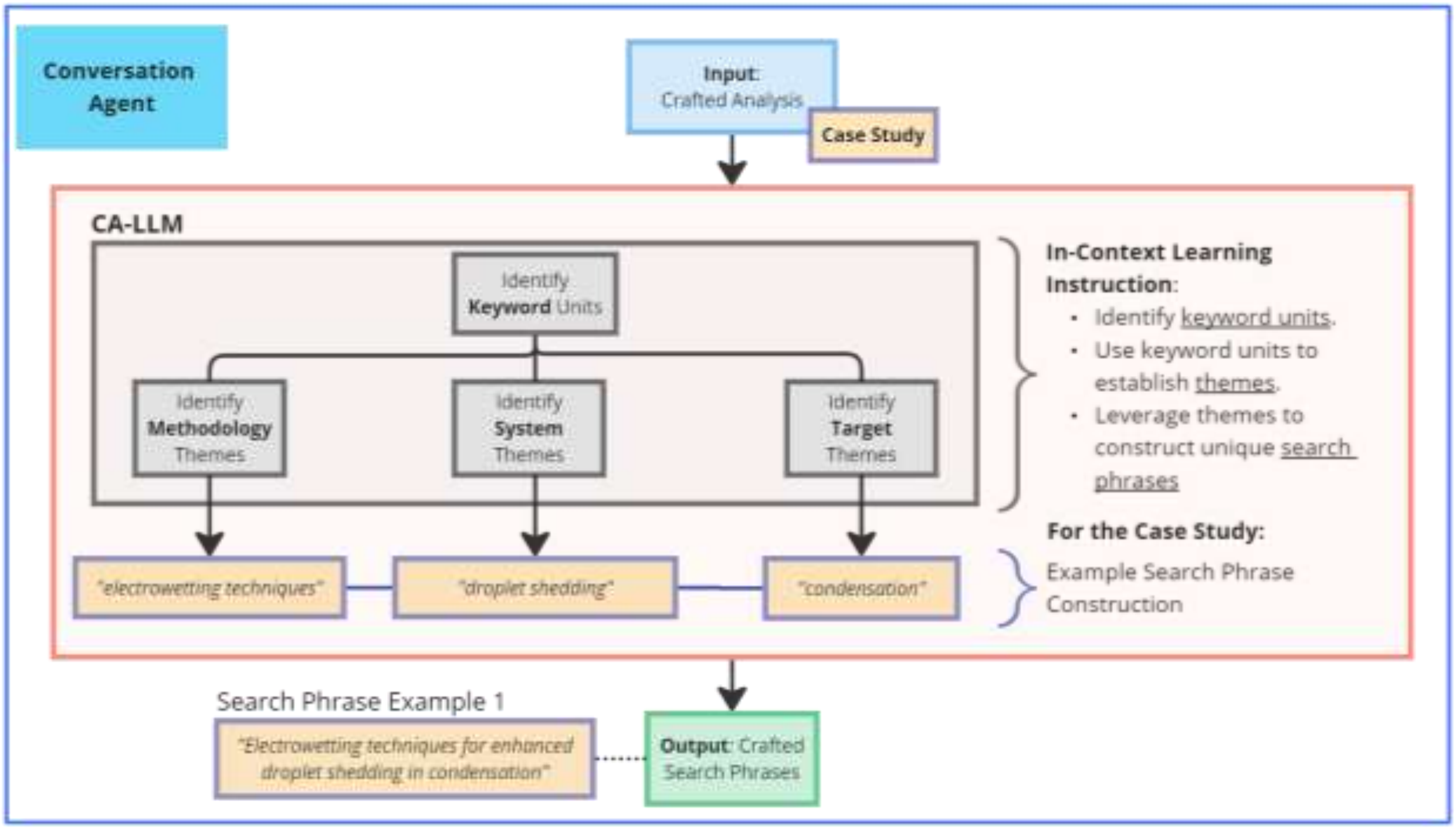

**Fig. S3. Natural language search phrase construction through a Conversation Agent LLM**. Displays the ICL instruction passed to the LLM, consisting of the bottom-up approach identifying keyword units, organizing them into themes, and finally creating unique search phrases. Additionally, the figure displays an example for the first crafted search phrase of the case study.

**Search Phrase Expansion**

- Search Phrase 1: "Electrowetting techniques for enhanced droplet shedding in condensation"
- Lateral Expansion 1: "Optimization of electrowetting procedures to improve droplet shedding efficiency in condensation systems"

- Search Phrase 2: "Design and arrangement of electrodes in electrowetting for condensation applications"
- Lateral Expansion 2: "Strategic positioning of conductive elements in electrowetting setups for efficient droplet shedding in condensation processes"

- Search Phrase 3: "Material properties influencing droplet shedding in condensation systems"
- Lateral Expansion 3: "Impact of surface characteristics on the removal of liquid droplets in condensation setups"

- Search Phrase 4: "Thermal conductivity and heat flux in dropwise condensation"
- Lateral Expansion 4: "Analysis of heat transfer rates and thermal properties in dropwise condensation mechanisms"

- Search Phrase 5: "The influence of contact angle and surface tension on droplet shedding in electrowetting applications"
- Lateral Expansion 5: "Relationship between liquid-solid interactions, surface properties, and droplet removal efficiency in electrowetting systems"

**Fig. S4. Lateral expansion results for the finalized user-selected phrases and custom input phrases.**

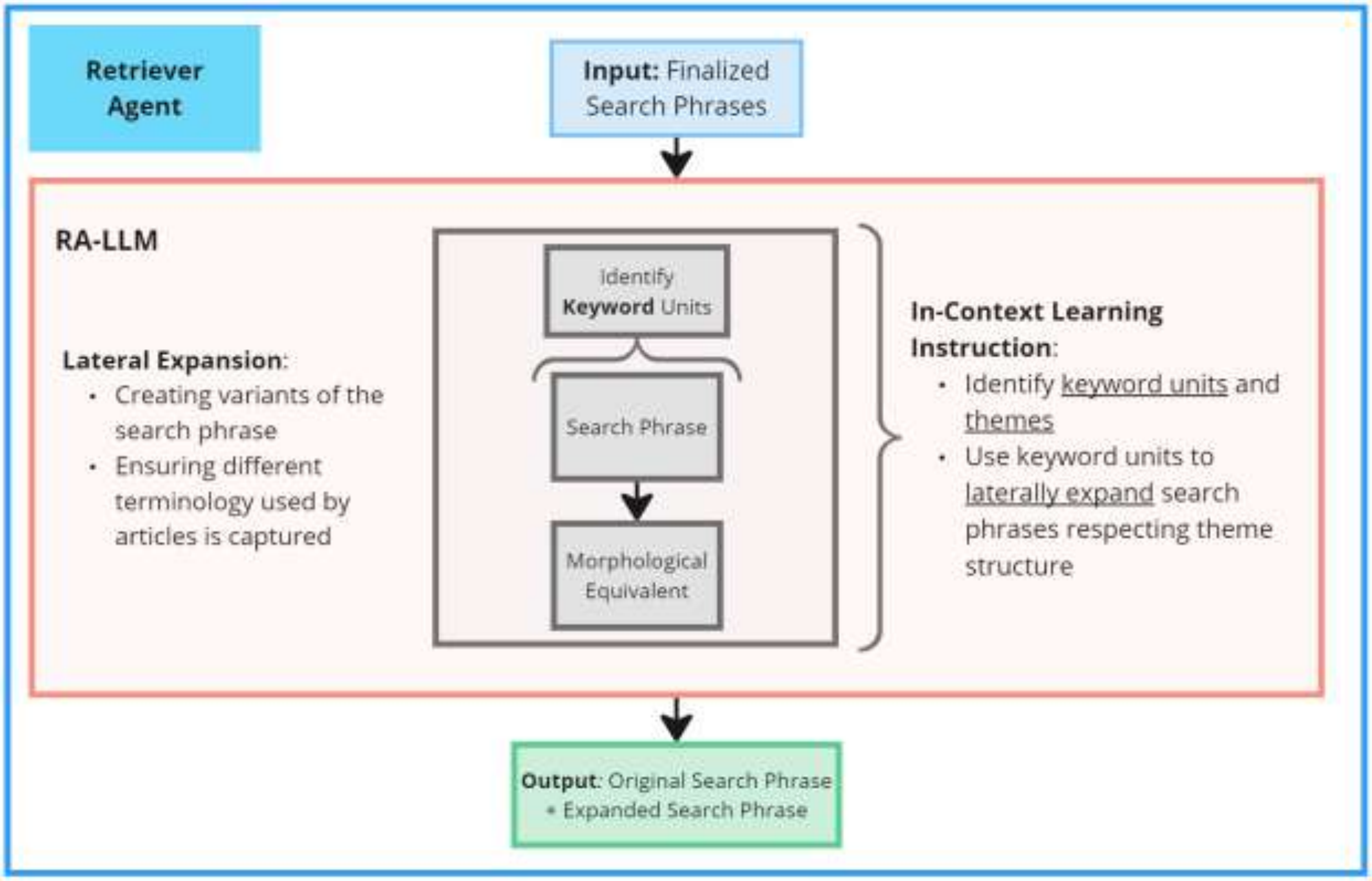

**Fig. S5. Lateral expansion process.** Displays the development of morphologically similar variants of the finalized phrases.

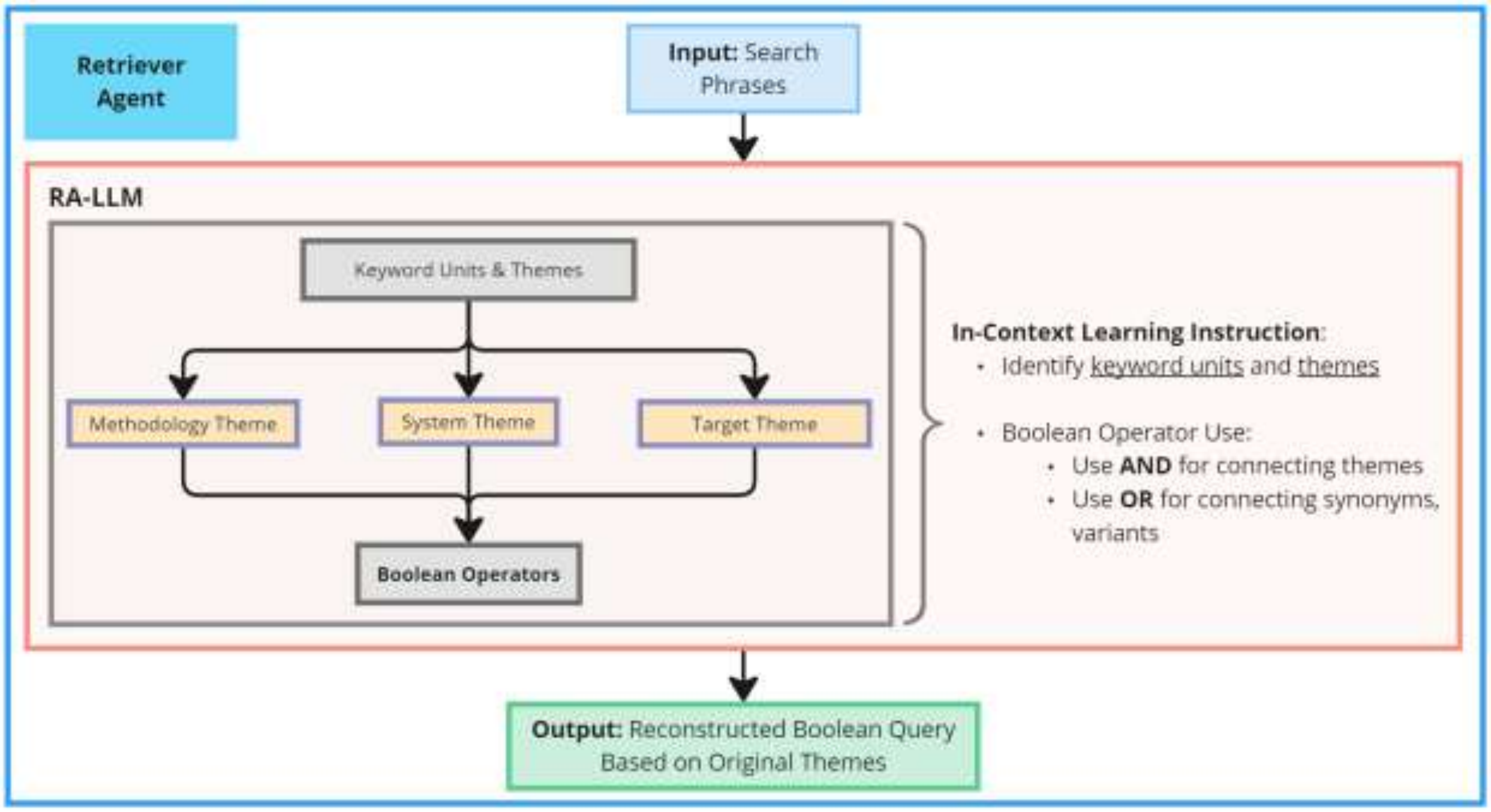

**Fig. S6. Boolean query transformation process.**

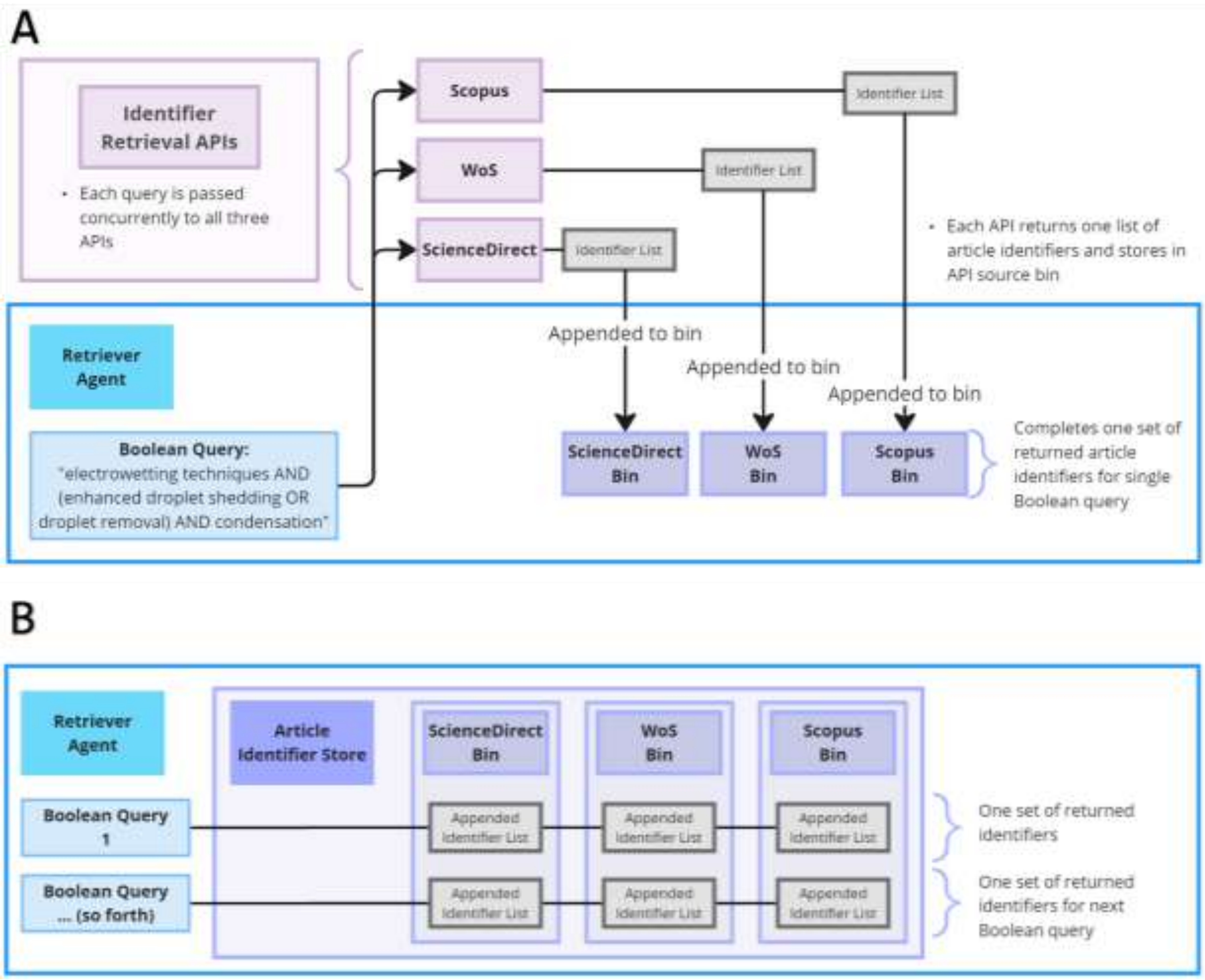

**Fig. S7. Article Identifier Retrieval Process.** (**A**) Example process for the article identifier retrieval of a case study query, demonstrating how returned identifiers are appended to their associated API bins. (**B**) Displays how each API bin will have multiple sets depending on how many Boolean queries were selected by the user.

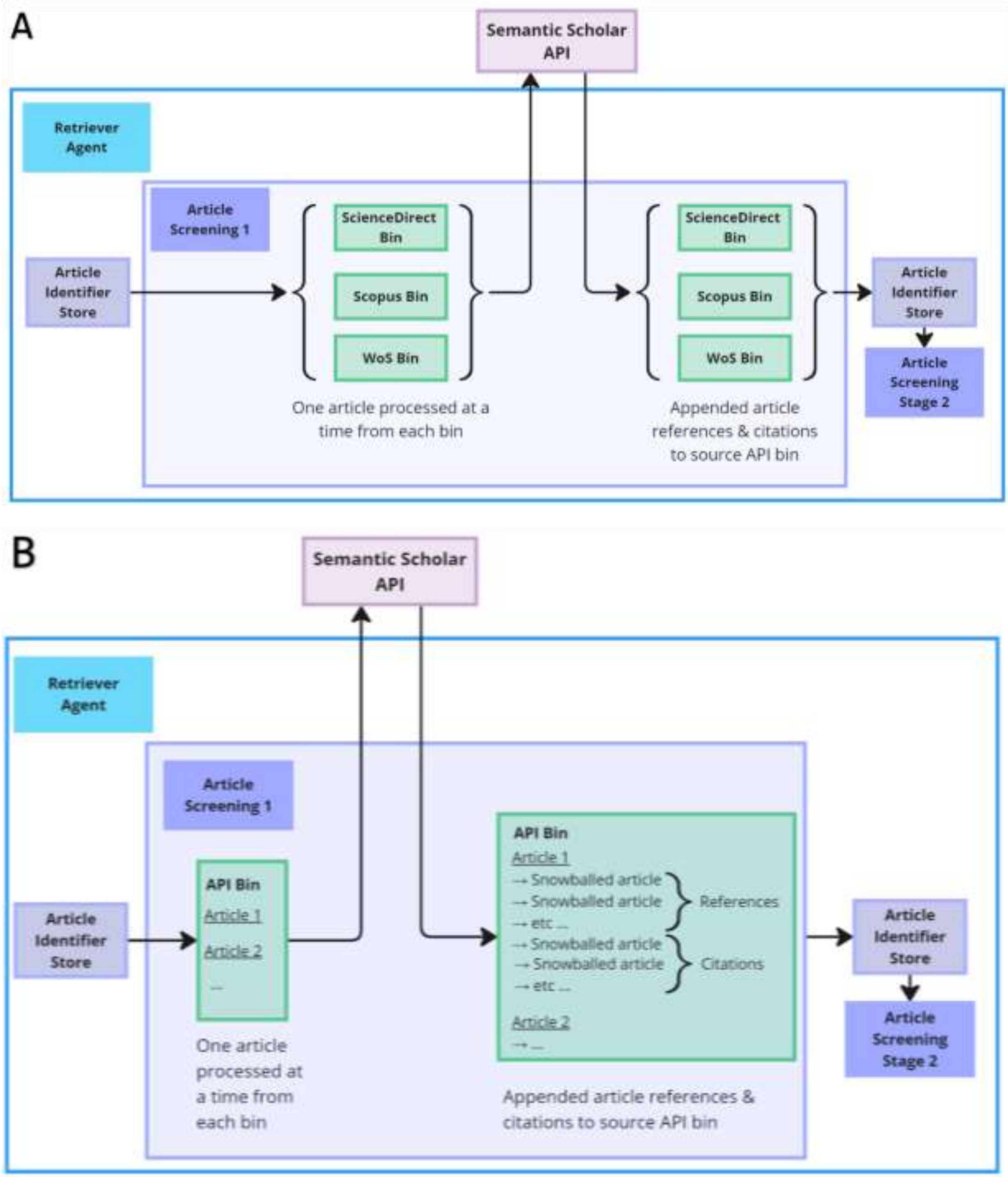

**Fig. S8. Snowballing screening process, leveraging Semantic API to track forward and backward references of article identifiers.** (**A**) a broader-level depiction for each article API bin and (**B**) a more detailed depiction through one API bin

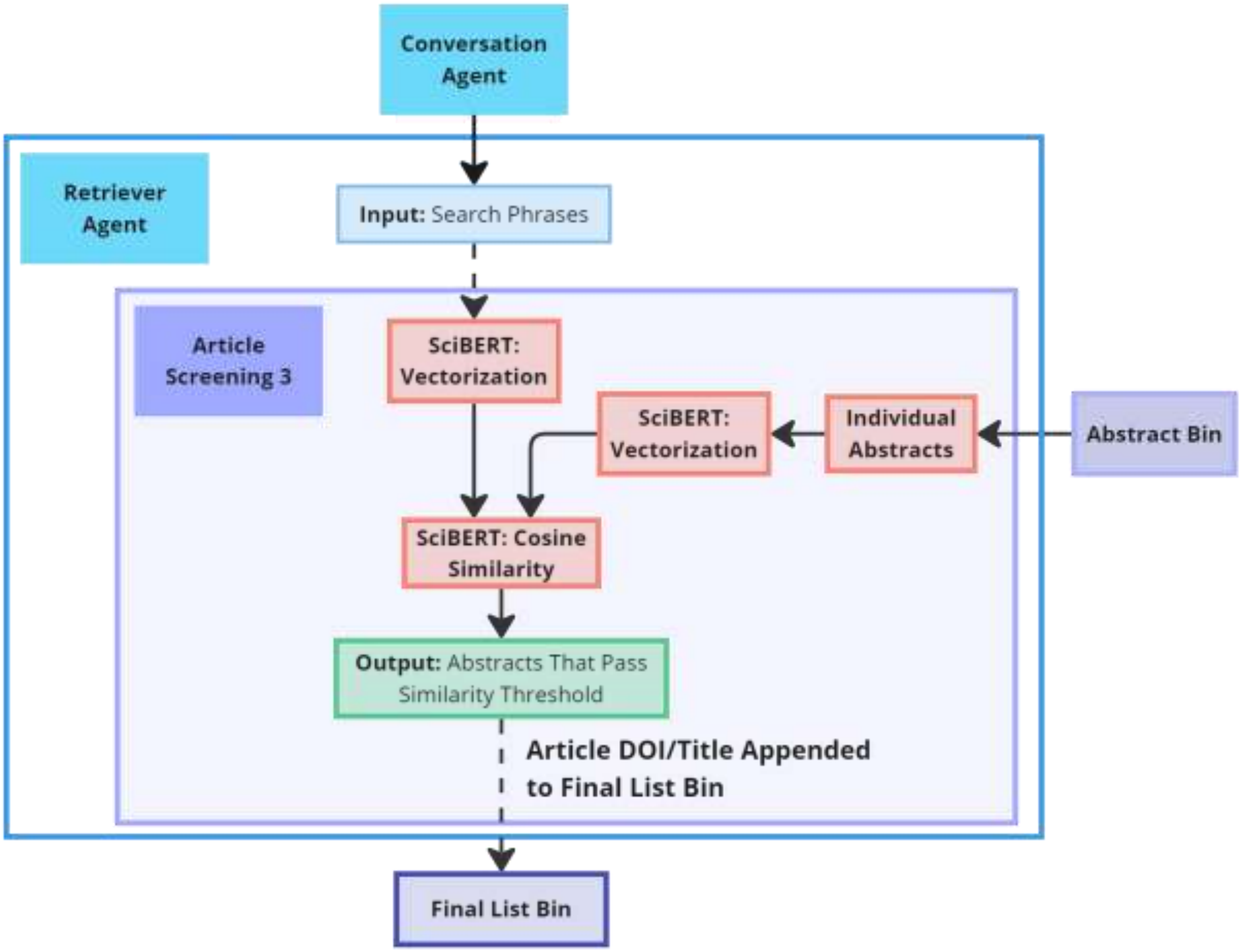

**Fig. S9. Semantic screening of low priority article identifiers.** Displays how Google's BERT LLM is utilized to identify higher ranked abstracts through cosine similarity computation.

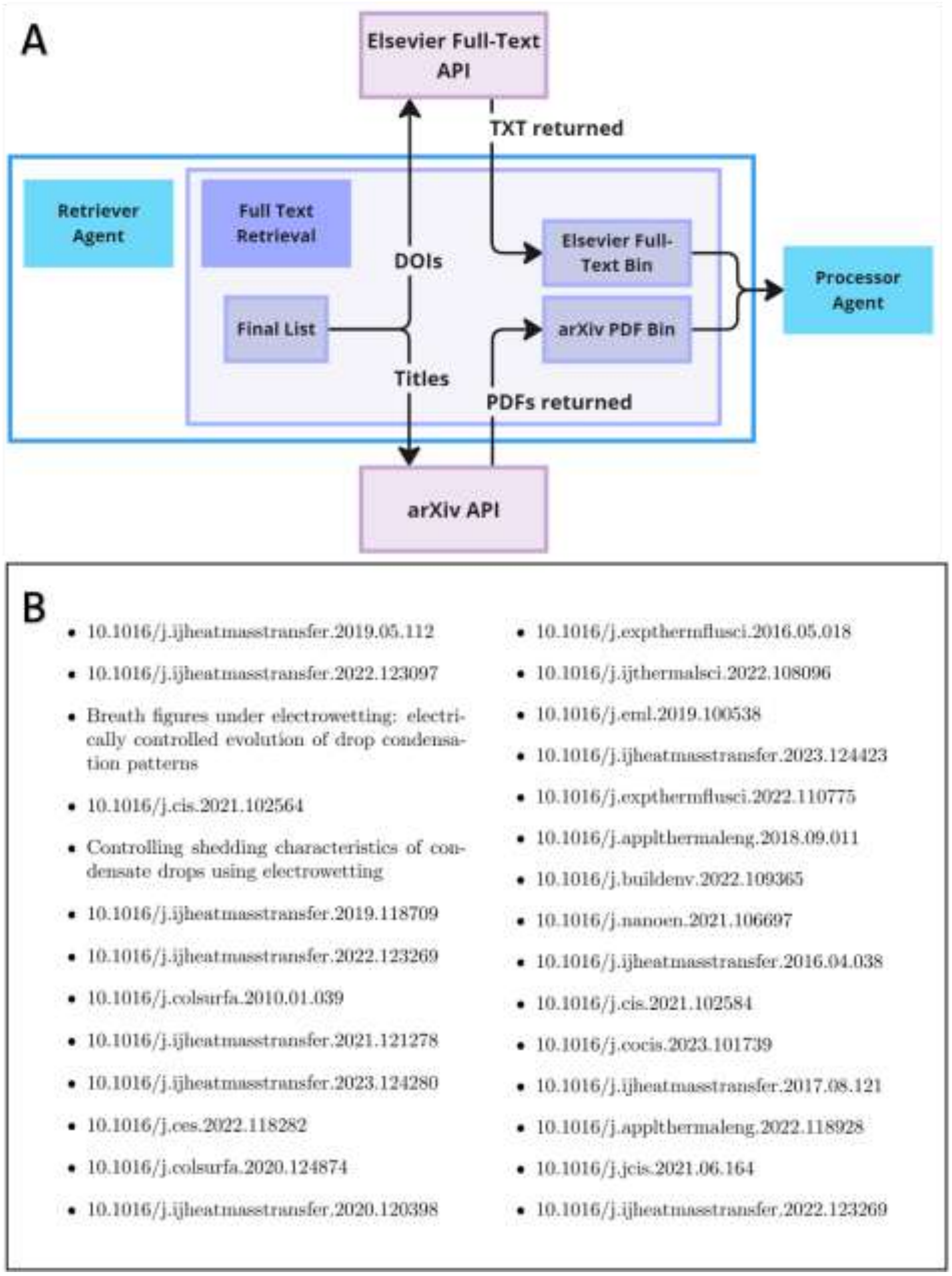

**Fig. S10. Full text retrieval of articles and results.** (**A**) Retrieval process for the article identifiers present on the Final List bin returning available full texts as PDFs or in TXT format. (**B**) Final 28 articles that contained available full text.

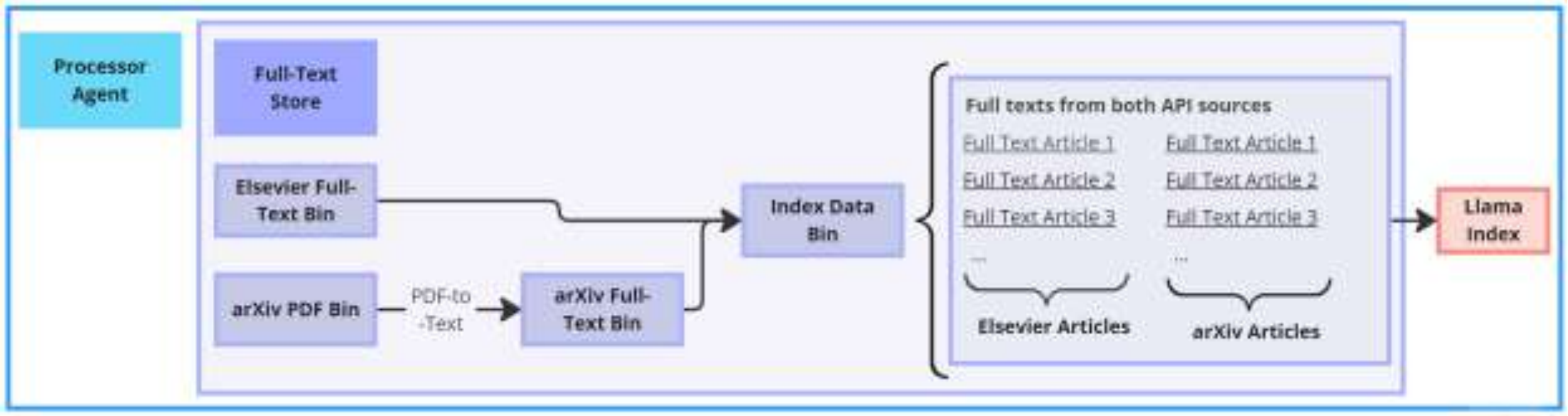

**Fig. S11. Full text article consolidation process.**

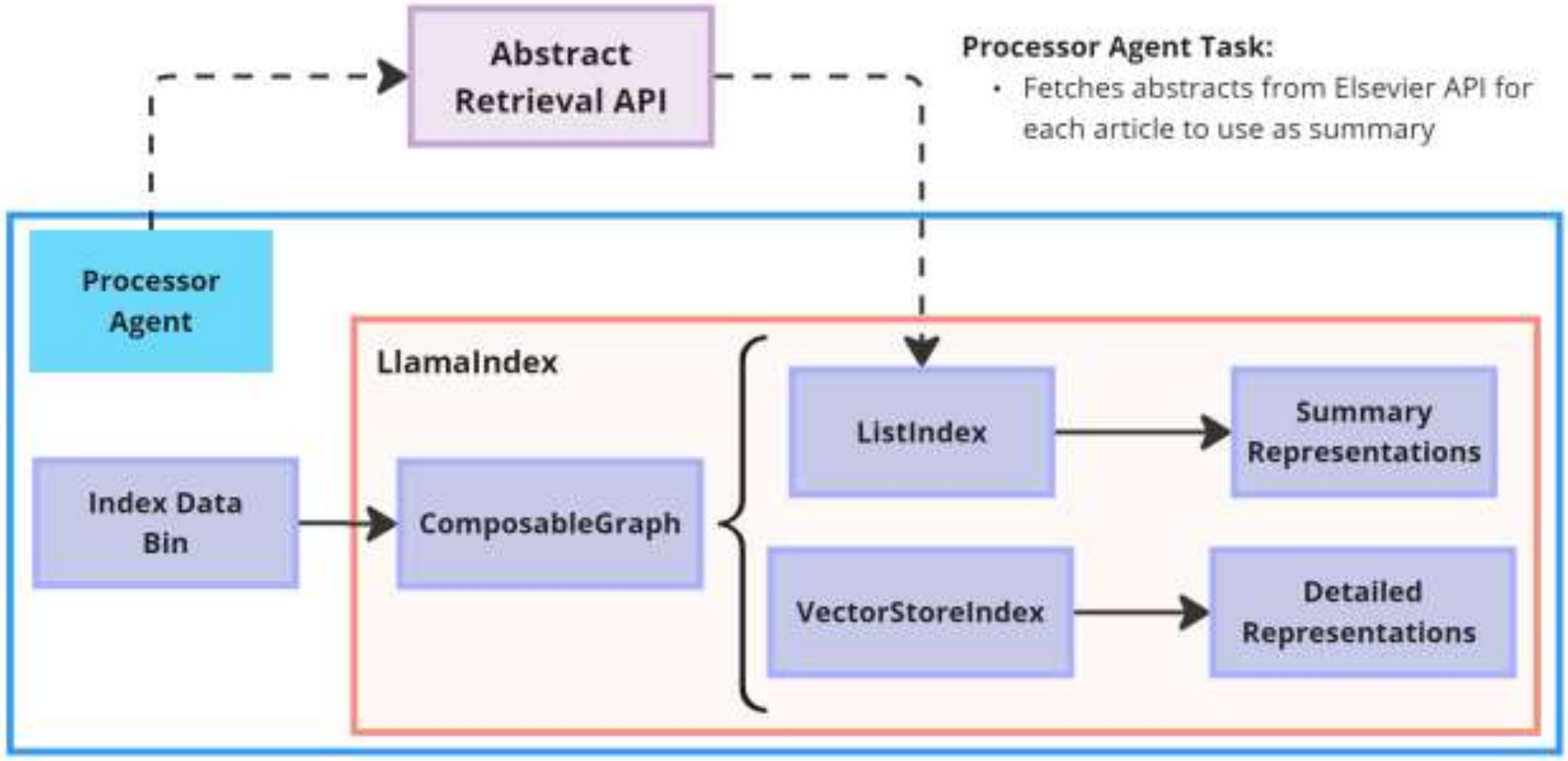

**Fig. S12. Depicts the creation of the composable graph structure through LlamaIndex.**

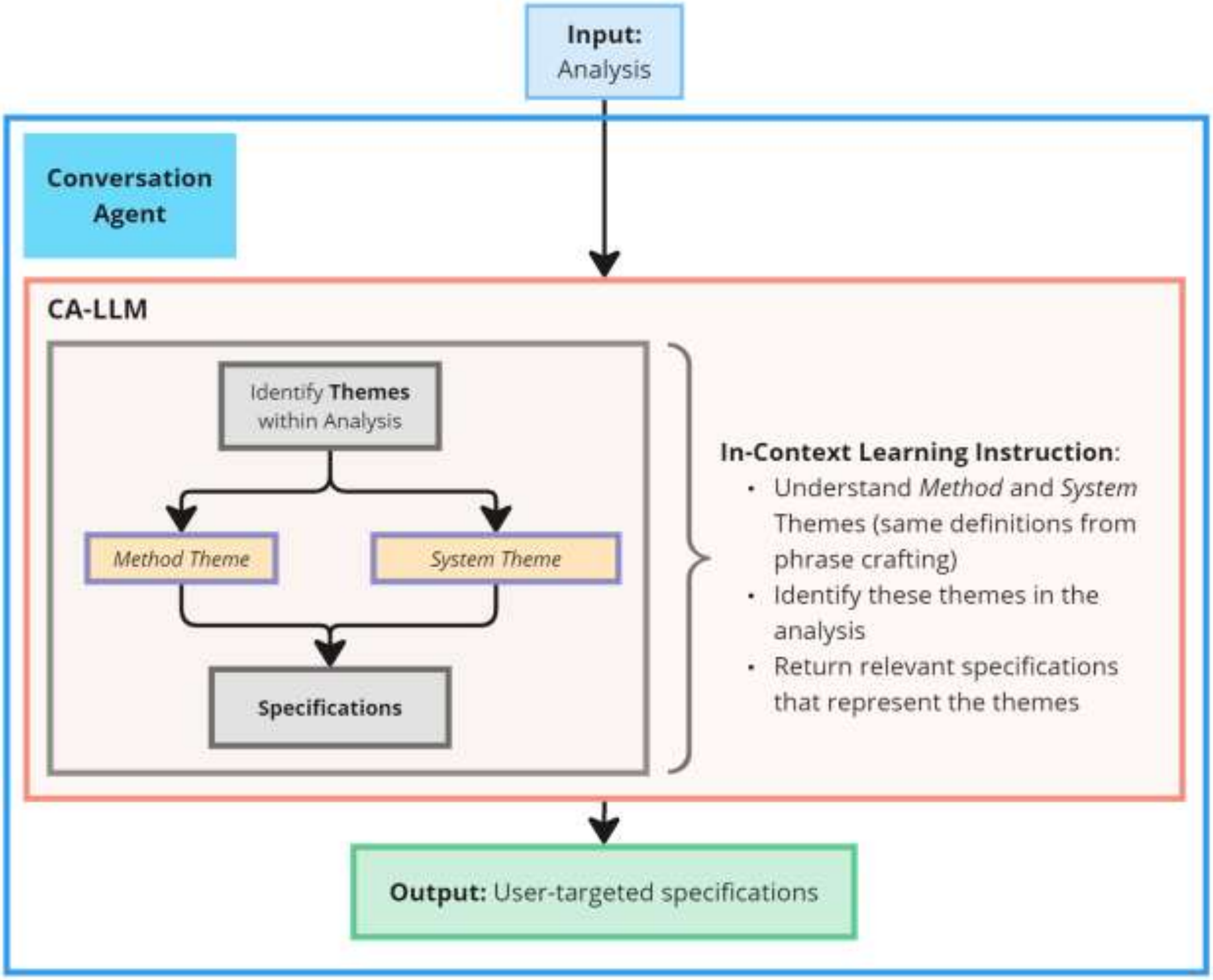

**Fig. S13. Conversation Agent user-targeted specification/parameter crafting process.** Displays how the LLM is programmed via ICL, to utilize the crafted analysis and previously identified *Themes* to craft specifications.

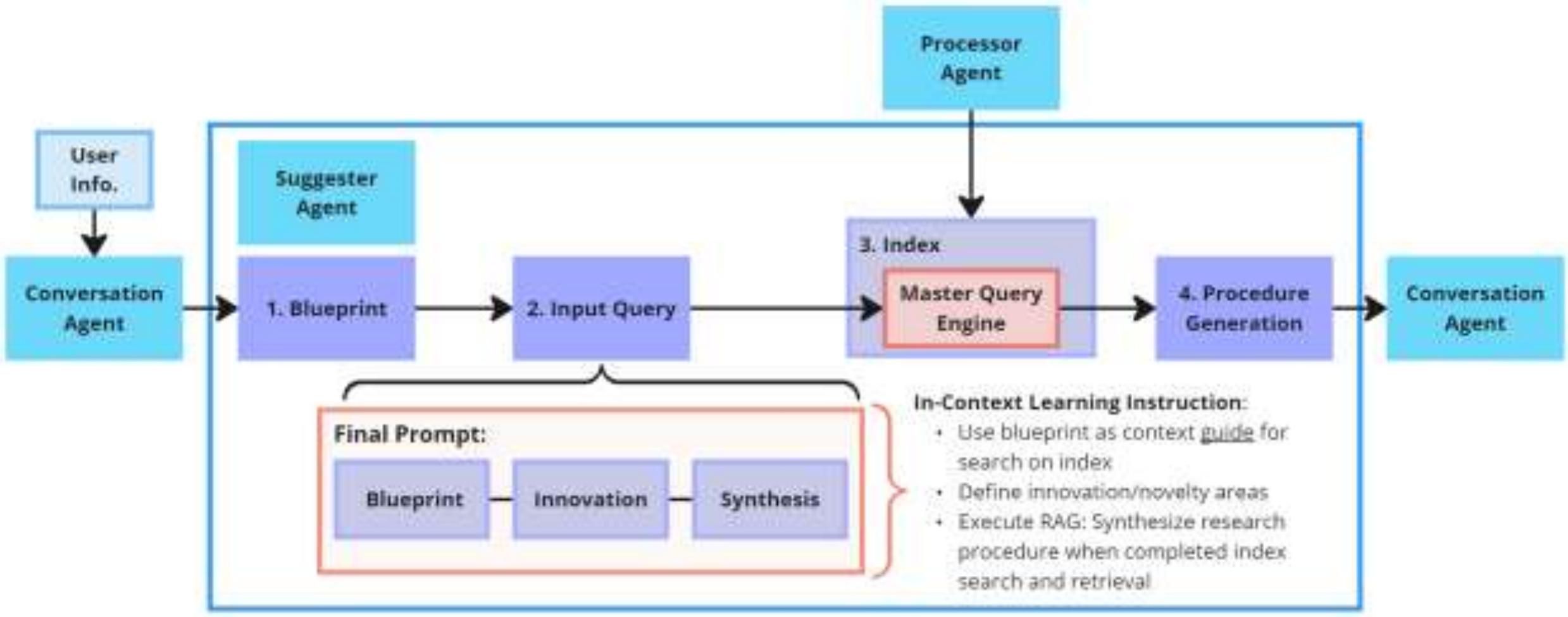

**Fig. S14. Suggester Agent final ICL prompt setup.** Displays how the input query, for RAG execution is conducted for procedure generation.

=== Re-Generated Research Procedure ===

Assistant:

**Experimental Procedure**

**1. Experimental Setup:**

a. Assemble a microfluidic device with a high surface area porous electrode. This device should have multiple microchannels with electrode connections at their beginnings and ends.

b. Embed the microchannels with thin plate electrodes of varying lengths and widths, with the ratio of the length of the high voltage electrode to the low voltage electrode being 1:3.

c. Embed the electrodes to the bottom heated wall to promote depletion of the thermal boundary layer and heat transfer.

**2. Materials and Equipment:**

a. Gather the following materials and equipment: a microfluidic device, a high voltage power supply, an optical read-out, polyester, cyclic olefin copolymer materials, and gold leaf.

b. Use de-ionized (DI) water as the working fluid.

c. Utilize a power supply to provide the necessary voltage to the electrodes, a thermocouple to measure the temperature of the fluid, and a flow meter to measure the flow rate of the fluid.

**3. Electrode Design and Arrangement:**

a. Optimize the electrode design and arrangement to achieve the desired droplet manipulation and shedding.

b. Adjust the design and arrangement of the electrodes to achieve the desired droplet speed and airflow velocity.

c. Design the electrodes to have a specific pattern, such as a periodic grid of hollow squares, with each electrode having a width of 50 $\mu$m and each square being 1000 $\mu$m in size.

**4. Droplet Generation and Manipulation:**

a. Generate the droplets by applying a voltage to the control electrodes.

b. Set the voltage frequency to 10 kHz and vary the voltage amplitude from 0 to 1 kV.

c. Vary the droplet size from 0.5 $\mu$L to 5$\mu$L.

d. Manipulate the droplets by applying a voltage to the control electrodes and adjust the voltage to achieve the desired droplet speed.

**5. Heat Transfer Measurement:**

a. Measure the heat transfer coefficient by monitoring the temperature of the droplet as it moves through the channel.

b. Use a thermocouple to measure the temperature.

c. Vary the heat transfer coefficient from 0.1 W/m2K to 10 W/m2K.

**6. Airflow Velocity Measurement:**

a. Measure the airflow velocity using a velocity meter.

b. Vary the airflow velocity from 0.1 m/s to 10 m/s.

7. Data Acquisition and Analysis:

a. Record the droplet movement using a high-speed camera and analyze the video using video analysis software.

b. Calculate the droplet velocity from the recorded video frames and plot the droplet trajectory.

c. Compare the droplet velocity with the theoretical model to evaluate the performance of the microfluidic device.

**8. Results:**

a. Analyze and discuss the results of the experiment in light of the research topic and queries.

b. Compare the accuracy of droplet shedding and heat transfer using the proposed device with other commercial devices.

c. Determine the effect of the electrode design and arrangement on the droplet manipulation and shedding.

d. Analyze the data to determine the effect of the droplet speed and airflow velocity on the heat transfer coefficient.

**9. Conclusion:**

a. Summarize the findings of the experiment.

b. Discuss the implications of the results for the research topic.

c. Explore potential for future research in the area.

**Fig. S15. Suggester Agent procedure re-generation.**